\documentclass[aps,prd,twocolumn,showpacs,amsmath,amssymb]{revtex4-2}
\usepackage{amsmath}
\usepackage{graphicx}
\usepackage{subfigure}
\usepackage{epstopdf}
\usepackage{color}
\usepackage{multirow}
\usepackage{setspace}
\usepackage{overpic}
\usepackage{amssymb}
\usepackage[bookmarksnumbered, pdfstartview=FitH,colorlinks,urlcolor=blue,
citecolor=blue,linkcolor=blue] {hyperref}
\usepackage{lineno}
\usepackage{bm}
\usepackage{rotating}
\usepackage[utf8]{inputenc}

\let\oldequation\equation \let\oldendequation\endequation
\renewenvironment{equation} {\linenomathNonumbers\oldequation}{\oldendequation\endlinenomath}
\begin{document}


\title{\bf \boldmath{ Measurement of the branching fractions of $D^+
    \to K^+K^-\pi^+\pi^+\pi^-$, $\phi\pi^+\pi^+\pi^-$, $K^0_SK^+\pi^+\pi^-\pi^0$,
$K^0_SK^+\eta$, and $K^0_SK^+\omega$ decays}}
\author{
M.~Ablikim$^{1}$, M.~N.~Achasov$^{4,c}$, P.~Adlarson$^{76}$, X.~C.~Ai$^{81}$, R.~Aliberti$^{35}$, A.~Amoroso$^{75A,75C}$, Q.~An$^{72,58,a}$, Y.~Bai$^{57}$, O.~Bakina$^{36}$, Y.~Ban$^{46,h}$
% ...
\\
\vspace{0.2cm}
(BESIII Collaboration)\\
\vspace{0.2cm} {\it
$^{1}$ Institute of High Energy Physics, Beijing 100049, People's Republic of China\\
% ...
$^{4}$ Budker Institute of Nuclear Physics SB RAS (BINP), Novosibirsk 630090, Russia\\
% ...
$^{35}$ Johannes Gutenberg University of Mainz, Johann-Joachim-Becher-Weg 45, D-55099 Mainz, Germany\\
$^{36}$ Joint Institute for Nuclear Research, 141980 Dubna, Moscow region, Russia\\
% ...
$^{46}$ Peking University, Beijing 100871, People's Republic of China\\
% ...
$^{57}$ Southeast University, Nanjing 211100, People's Republic of China\\
$^{58}$ State Key Laboratory of Particle Detection and Electronics, Beijing 100049, Hefei 230026, People's Republic of China\\
% ...
$^{72}$ University of Science and Technology of China, Hefei 230026, People's Republic of China\\
% ...
$^{75}$ University of Turin and INFN, (A)University of Turin, I-10125, Turin, Italy; (B)University of Eastern Piedmont, I-15121, Alessandria, Italy; (C)INFN, I-10125, Turin, Italy\\
$^{76}$ Uppsala University, Box 516, SE-75120 Uppsala, Sweden\\
% ...
$^{81}$ Zhengzhou University, Zhengzhou 450001, People's Republic of China\\
\vspace{0.2cm}
$^{a}$ Deceased\\
% ...
$^{c}$ Also at the Novosibirsk State University, Novosibirsk, 630090, Russia\\
% ...
$^{h}$ Also at State Key Laboratory of Nuclear Physics and Technology, Peking University, Beijing 100871, People's Republic of China\\
% ...
}
}

\begin{abstract}
  Using $20.3~\mathrm{fb}^{-1}$ of $e^+e^-$ collision data
  collected at a center-of-mass energy of 3.773~GeV with the BESIII
  detector operating at the BEPCII collider, the branching fractions of three hadronic charm meson decays,
  $D^+\to \phi\pi^+\pi^+\pi^-$, $D^+\to K^0_SK^+\pi^+\pi^-\pi^0$, and $D^+\to
  K^0_SK^+\omega$, are measured for the first time to be
  $(0.54\pm0.19\pm0.02)\times 10^{-4}$, $(2.51\pm0.34\pm0.14)\times 10^{-4}$, and
  $(2.02\pm0.35\pm0.10)\times 10^{-4}$, respectively.  Futhermore, the branching fractions of
  $D^+\to K^+K^-\pi^+\pi^+\pi^-$ and $D^+\to K^0_SK^+\eta$ are measured with
  improved precision, yielding values of $(0.66\pm0.11\pm0.03)\times 10^{-4}$ and
  $(2.27\pm0.22\pm0.05)\times 10^{-4}$, respectively.
\end{abstract}

\maketitle

\section{Introduction}

Hadronic $D$ decays provide an ideal platform for exploring strong and weak
interactions in decays of hadrons with charm or bottom quarks~\cite{Li:2021iwf}.
In recent years, the branching fraction
(BF) ratios ${\mathcal R}_{\tau/\ell}={\mathcal B}_{B\to \bar
  D^{(*)}\tau^+{\bar \nu}_\tau}/{\mathcal B}_{B\to \bar D^{(*)}\ell^+{\bar
\nu}_\ell}$~($\ell=\mu$, $e$) measured by
different experiments,
have been found to deviate from the standard model (SM) prediction by
$\approx 3.1\sigma$~\cite{hflav2018},
and the exclusive hadronic $D^{0(+)}$ decays with three charged pions in the final
state are key potential backgrounds in this measurement~\cite{lhcbnote}.
To date, the BFs of many
exclusive hadronic $D^{0(+)}$ decays are poorly studied~\cite{ref::pdg2024}.
Measurements of the BFs of these decays
offer important input for
lepton flavor universality tests with semileptonic $B$ decays,
and studies of multi-body charm decays also help model and understand
the background when probing the $D^{(*)}$ decay with a $\tau$ in the final state.

Although many studies of hadronic $D$ decays have been performed by different experiments~\cite{ref::pdg2024},
knowledge of the hadronic decays $D^{+}\to
K\bar K \pi\pi\pi$ remains limited. Previously, only
the BF of the decay $D^+\to K^+K^-\pi^+\pi^+\pi^-$ was measured by the FOCUS
Collaboration, yielding $(2.3 \pm 1.2) \times 10^{-4}$~\cite{kkppp}, and the BF of the decay
$D^+\to K^0_SK^+\eta$ was measured by the BESIII Collaboration, yielding $(1.8 \pm
0.5) \times 10^{-4}$~\cite{kketa}.  In this work, the measurements
of the absolute BFs of the decays $D^+\to K^+K^-\pi^+\pi^+\pi^-$, $D^+\to
K^0_SK^+\pi^+\pi^-\pi^0$, $D^+\to \phi\pi^+\pi^+\pi^-$, $D^+\to K^0_SK^+\eta$,
and $D^+\to K^0_SK^+\omega$ are performed by analyzing $20.3~\mathrm{fb}^{-1}$ of $e^+e^-$ collision data
collected at the center-of-mass energy $\sqrt s=3.773$~GeV with the BESIII
detector~\cite{BESIII:2024lbn}. Throughout this paper, charge conjugation is always implied.

\section{BESIII detector and Monte Carlo simulation}

The BESIII detector~\cite{Ablikim:2009aa} records symmetric $e^+e^-$ collisions
provided by the BEPCII storage ring~\cite{Yu:IPAC2016-TUYA01} in the
center-of-mass energy range from 1.84 to 4.95~GeV, with a peak luminosity of
$1.1 \times 10^{33}\;\text{cm}^{-2}\text{s}^{-1}$ achieved at $\sqrt{s} =
3.773\;\text{GeV}$.  BESIII has collected large data samples in this energy
region~\cite{Ablikim:2019hff,EcmsMea,EventFilter}. The cylindrical core of the
BESIII detector~\cite{detvis} covers 93\% of the full solid angle and consists
of a helium-based multilayer drift chamber~(MDC), a plastic scintillator
time-of-flight system~(TOF), and a CsI(Tl) electromagnetic calorimeter~(EMC),
which are all enclosed in a superconducting solenoidal magnet providing a 1.0~T
magnetic field.

The solenoid is supported by an octagonal flux-return yoke with resistive plate
counter muon identification modules interleaved with steel.  The
charged-particle momentum resolution at $1~{\rm GeV}/c$ is $0.5\%$, and the
$dE/ dx$ resolution is $6\%$ for electrons from Bhabha scattering. The EMC
measures photon energies with a resolution of $2.5\%$ ($5\%$) at $1$~GeV in the
barrel (end-cap) region. The time resolution in the TOF barrel region is 68~ps,
while that in the end-cap region was 110~ps.  The end-cap TOF system was
upgraded in 2015 using multigap resistive plate chamber technology, providing a
time resolution of 60~ps, which benefits 86\% of the data used in this
analysis~\cite{etof}.

Simulated Monte Carlo (MC) samples produced with the {\sc geant4}-based~\cite{geant4}
MC package which includes the geometric description of the BESIII
detector and the detector response, are used to determine the detection
efficiency and estimate the backgrounds. The simulation includes the
beam-energy spread and initial-state radiation in the $e^+e^-$ collisions
modeled with the {\sc kkmc} generator~\cite{kkmc}.  An inclusive MC sample
includes the production of $D\bar{D}$ pairs (including quantum coherence for
the neutral $D$ channels), the non-$D\bar{D}$ decays of the $\psi(3770)$, the
initial-state radiation production of the $J/\psi$ and $\psi(3686)$ states, and
the continuum processes incorporated in {\sc kkmc}~\cite{kkmc}.  All particle
decays are modelled with {\sc evtgen}~\cite{evtgen} using the BFs either taken
from the Particle Data Group~(PDG)~\cite{ref::pdg2024}, when available, or otherwise
estimated with {\sc lundcharm}~\cite{lundcharm}. Final-state radiation from
charged particles is incorporated with the {\sc photos} package~\cite{photos}.

The signal MC samples for the decays $D^+\to K^{+}K^{-}\pi^+\pi^+\pi^- $,
$D^+\to K^0_SK^+\pi^+\pi^-\pi^0$ and $D^+\to K^0_S K^+\omega$
are generated with potential resonances, using the fractions listed in Table~\ref{mix},
while the samples for other signal decays are generated using a uniform phase space.

\begin{table}[htbp] \centering \caption{The fractions of various
    resonance states for the decays $D^+\to K^+K^-\pi^+\pi^+\pi^-$, $D^+\to K^0_SK^+\pi^+\pi^-\pi^0,$ and $D^+\to K^0_SK^+\omega$.
    Except for those listed below, no other significant resonances are observed in the data due to limited sample sizes,
    and the fractions for the former two signal decays are obtained based on our measurements later.
  } \label{mix} \centering
  \resizebox{\linewidth}{!}{
    \begin{tabular}{|l|c|c|c|c|c|}
      \hline\hline Signal decay                          & Resonance source
                                                         & Fraction (\%) \\ \hline
      \multirow{2}*{$D^+\to K^+K^-\pi^+\pi^+\pi^-$}   &
      $K^+K^-\pi^+\pi^+\pi^-$            & $60.0  \pm 14.0 $  \\
      \cline{2-3} & $\phi\pi^+\pi^+\pi^-$                    & $40.0 \pm 14.0 $
      \\ \cline{2-3} \cline{1-3}
      \multirow{2}*{$D^+\to K^0_SK^+\pi^+\pi^-\pi^0$} &
      $K^0_S K^+ \eta$                         & $26.9 \pm 9.2 $  \\ \cline{2-3}
                                               & $ K^0_S K^+ \omega$                      & $73.1 \pm 9.2 $  \\ \hline
      \multirow{2}*{$D^+\to K^0_SK^+\omega$}          & $\overline{K}_{1}^{0} K^+ $ & $56.2 \pm 2.4 $~\cite{FOCUS:2001omf}  \\ \cline{2-3}
                                                      & $K_{1}^{+} K^{0}_{S} $       & $43.8 \pm 2.4
                                                      $~\cite{FOCUS:2001omf}  \\ \hline\hline
    \end{tabular}
  }
\end{table}

\section{Method} \label{sec:method}

The $\psi(3770)$ resonance, which lies just above the $D\bar D$
threshold, predominantly decays into $D^0\bar D^0$ and $D^+D^-$ meson pairs. Taking this
advantage, the absolute BF of the $D^+$ decay can be measured using the double-tag
(DT) method, which was first developed by the MARKIII
Collaboration~\cite{MARKIII1,MARKIII2}. To measure the BF of a given $D^+$
decay, single-tag (ST) $D^-$ mesons are selected using the six hadronic
decay modes $K^{+}\pi^{-}\pi^{-}$, $K^{0}_{S}\pi^{-}$,
$K^{+}\pi^{-}\pi^{-}\pi^{0}$, $K^{0}_{S}\pi^{-}\pi^{0}$,
$K^{0}_{S}\pi^{+}\pi^{-}\pi^{-}$, and  $K^{+}K^{-}\pi^{-}$, which have
relatively large BFs and low backgrounds.  Events in which a signal candidate is
reconstructed in the presence of an ST $D^-$ meson are referred to as DT
events. The BF of the signal decay can be determined as
\begin{equation}
  {\mathcal B}_{\rm sig} = \frac{N_{\rm DT}}{N^{\rm tot}_{\rm
  ST}\cdot \epsilon_{{\rm sig}}},
\end{equation}
where $N^{\rm tot}_{\rm ST} = \sum_{i}N^{i}_{\rm ST}$ and $N_{\rm DT}$ represent
the total yields of the ST and DT candidates in the data, respectively, and
$N_{\rm ST}^{i}$ is the ST yield for tag mode $i$.
The efficiency for detecting the signal $D^{+}$ decay, averaged over all tag modes, is given by
\begin{equation}
  \epsilon_{{\rm sig}} = \sum_{i} \frac{\epsilon^{i}_{\rm DT}}{\epsilon^{i}_{\rm ST}} \cdot \frac{N^{i}_{\rm ST}}{N^{\rm tot}_{\rm ST} },
\end{equation}
where $\epsilon^{i}_{\rm ST}$ is the efficiency of
reconstructing the ST mode $i$, and
$\epsilon^{i}_{\rm DT}$ is the efficiency of
finding both the ST mode $i$ and the signal decay simultaneously.
\section{Single-tag selection}

For each charged track (except those used for $K^0_S$ reconstruction), the
polar angle with respect to the MDC $z$-axis ($\theta$) satisfies
$|\cos\theta|<0.93$, and the point of closest approach to the interaction
point~(IP) must be within 1\,cm in the plane perpendicular to the $z$-axis and
within 10\,cm along the $z$-axis.
For particle identification (PID), charged tracks are identified using the
$dE/dx$ and TOF information, from which the combined confidence levels for
the pion and kaon hypotheses are computed separately. The charged tracks are
then assigned as the particle type with the higher probability.

The $K_S^0$ candidates are formed from pairs of oppositely charged tracks. For
these two tracks, the distance of closest approach to the IP is required to be
less than 20\,cm along the $z$-axis. There are no limitations on the distance
of closest approach in the transverse plane or on the PID criteria for these tracks. The two charged tracks are constrained to originate
from a common vertex, which must be at least twice the vertex resolution away
from the IP in terms of flight distance. The quality of the vertex fits
(primary vertex fit and secondary vertex fit) is ensured by a requirement on the
$\chi^2$ ($\chi^2 < 100$).  The invariant mass of the $\pi^+\pi^-$ pair is
required to be within $(0.487,0.511)$~GeV/$c^2$.

The $\pi^0$ candidates are reconstructed via $\pi^0 \to\gamma\gamma$ decays.
Photon candidates are selected from the EMC showers that are unassociated with any charged track~\cite{emcshower}, and
the EMC time deviation from the event start time is required to be within
[0,\,700]\,ns.  The energy deposited in the EMC  is required to be greater than
25 MeV in the barrel region ($|\cos\theta|<0.80$) and 50~MeV in the end-cap
region ($0.86<|\cos\theta|<0.92$).  The opening angle between the photon
candidate and the nearest charged track in the EMC is required to be greater
than $10^{\circ}$. At least one of the photons is required to be detected in
the barrel EMC, as the endcaps have poorer resolution. For each
$\pi^0$ candidate, the invariant mass of the photon pair is required to be
within $(0.115,\,0.150)$~GeV$/c^{2}$. To improve the momentum resolution, the
$\gamma\gamma $ invariant mass is constrained to the nominal $\pi^0$
mass~\cite{ref::pdg2024} using a 1-C kinematic fit, with the $\chi^{2}$ of the fit required to be less than 50.
The four-momentum of the $\pi^0$ candidate,
updated by this fit, is then used for further analysis.

To distinguish the ST $D^-$ mesons from combinatorial backgrounds,
we define the energy difference $\Delta E\equiv E_{D^-}-E_{\mathrm{beam}}$ and the beam-constrained mass $M_{\rm
BC}\equiv\sqrt{E_{\mathrm{beam}}^{2}/c^{4}-|\vec{p}_{D^-}|^{2}/c^{2}}$, where $E_{\mathrm{beam}}$ is the beam energy, and $E_{D^-}$ and $\vec{p}_{D^-}$ represent the total energy and momentum of the ST $D^-$ meson in the $e^+e^-$ center-of-mass frame.
If there is more than one $D^-$ candidate for each ST
mode, the one with the smallest $|\Delta E|$ is retained for further analyses.
The $\Delta E$ requirements and ST efficiencies are summarized in
Table~\ref{ST:realdata}.

\begin{figure*}[htbp]
  \vspace{-0.cm}
  \hspace{-0.cm}
  \scalebox{1.225}{
    \hspace{-0.75cm}
    \includegraphics[width=0.9\linewidth]{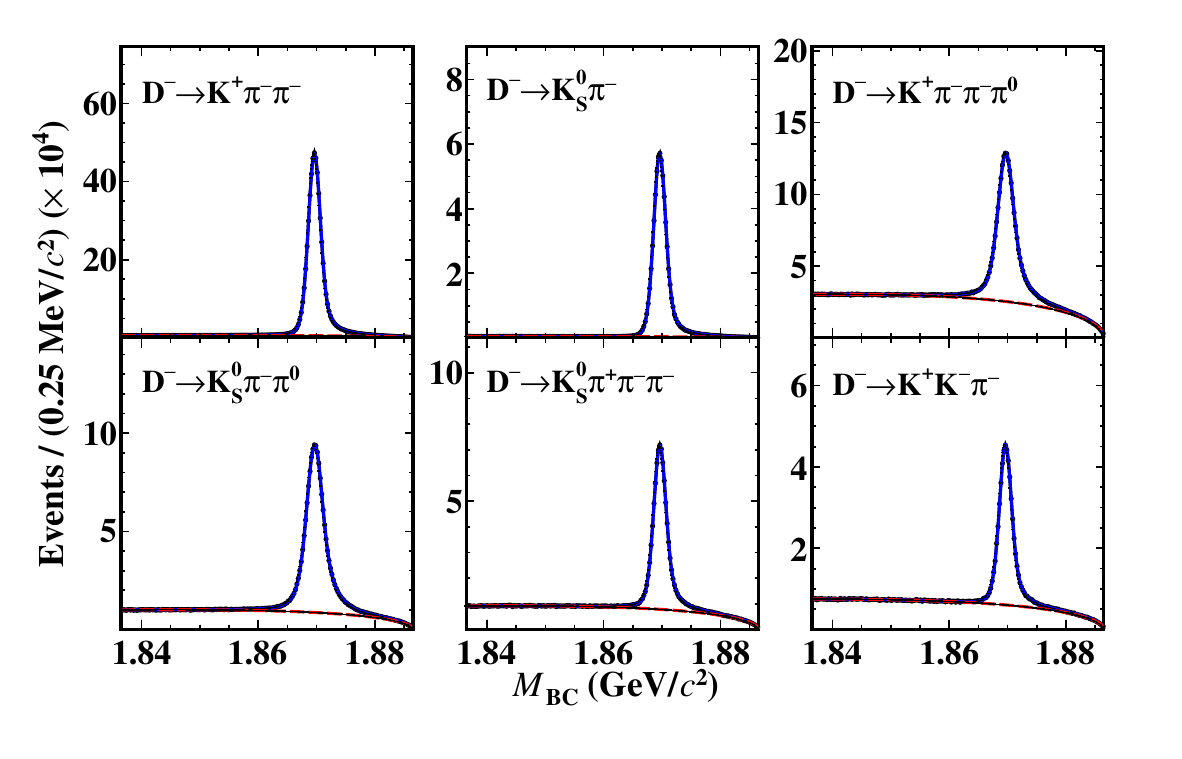}
  }
  \caption{
    The $M_{\rm BC}$ distributions of the ST $D^-$ candidates, with fit results overlaid.
    The points with error bars are data, the blue curves correspond to the best fit results, and
  the black solid curve with red dashes represents the fitted combinatorial background shapes.}
  \label{fig:datafit_Massbc}
\end{figure*}

For each tag mode, the yield of ST $D^-$ mesons is extracted by fitting the corresponding $M_{\rm BC}$ distribution.
The signal shape in the fit is modeled by the MC-simulated signal shape, convolved with a double-Gaussian function.
The background shape is described by an ARGUS function~\cite{argus}, with the endpoint fixed at 1.8865~GeV/$c^{2}$, corresponding to $E_{\rm beam}$.
Figure~\ref{fig:datafit_Massbc} shows the results of the fits to the $M_{\rm BC}$ distributions of the accepted ST candidates in the data for various tag modes.

\begin{table*}
  \renewcommand{\arraystretch}{1.2}
  \centering
  \caption {The $\Delta E$ requirements, ST yields in data~($N_{\rm ST}$), ST efficiencies~($\epsilon_{\rm ST}$) and DT efficiencies~($\epsilon_{\rm DT}$),
    where the uncertainties are statistical only.
    The superscripts 1, 2, 3, 4, 5 and 6 correspond to the following decays: $D^+\to K^+ K^- \pi^+ \pi^+ \pi^-$,
    $D^+\to K_{S}^{0} K^+ \pi^+ \pi^- \pi^0$,
    $D^+\to K_{S}^{0} K^+ \eta_{\gamma\gamma}$,
    $D^+\to \phi \pi^+ \pi^+ \pi^-$,
    $D^+\to K_{S}^{0} K^+ \eta_{\pi^+ \pi^- \pi^0}$, and
  $D^+\to K_{S}^{0} K^+ \omega$, respectively.}
  \scalebox{0.89}{
    \begin{tabular}{cccccccccc} \hline \hline Tag mode & $\Delta E$~(MeV)     &  $N_{\rm ST}~(\times 10^3)$   & $\epsilon_{\rm ST}$~(\%) & $\epsilon^{1}_{\rm DT}$~(\%)& $\epsilon^{2}_{\rm DT}$~(\%)& $\epsilon^{3}_{\rm DT}$~(\%)& $\epsilon^{4}_{\rm DT}$~(\%)& $\epsilon^{5}_{\rm DT}$~(\%)& $\epsilon^{6}_{\rm DT}$~(\%) \\ \hline
      $D^{-} \to K^{+}\pi^{-}\pi^{-}$       &   $(-25,24)$ &  $5710.4\pm2.5$ & $52.44\pm0.01$   &$4.72\pm0.06$&$2.33\pm0.05$&$16.52\pm0.16$&$6.39\pm0.10$  &$6.81\pm0.11$&$3.25\pm0.08$ \\
      $D^{-} \to K_{S}^{0}\pi^{-}$   & $(-25,26)$ & $~666.9\pm0.8$ &  $51.89\pm0.02$   &$4.82\pm0.06$ &$2.43\pm0.05$&$16.75\pm0.16$&$6.38\pm0.10$&$6.97\pm0.11$&$3.36\pm0.08$  \\
      $D^{-} \to K^{+}\pi^{-}\pi^{-}\pi^{0}$   & $(-57,46)$ & $1810.0\pm1.9$ & $27.19\pm0.01$ & $1.99\pm0.06$ &$0.82\pm0.04$&$13.35\pm0.29$&$5.57\pm0.20$ &$4.80\pm0.18$&$2.20\pm0.13$   \\
      $D^{-} \to K_{S}^{0}\pi^{-}\pi^{0}$        & $(-62,49)$ & $1507.2\pm1.5$ & $27.57\pm0.01$ &$2.19\pm0.06$ &$1.01\pm0.04$&$14.37\pm0.28$&$5.68\pm0.19$  &$5.54\pm0.18$&$2.55\pm0.13$   \\
      $D^{-} \to K_{S}^{0}\pi^{+}\pi^{-}\pi^{-}$ & $(-28,27)$ & $~809.7\pm1.1$ & $29.68\pm0.01$ & $2.27\pm0.06$ &$1.06\pm0.04$&$14.40\pm0.26$&$5.52\pm0.17$&$5.15\pm0.16$&$2.56\pm0.12$  \\
      $D^{-} \to  K^{+}K^{-}\pi^{-}$     & $(-24,23)$ & $~491.9\pm0.9$ & $42.05\pm0.02$   &$3.75\pm0.06$&$1.75\pm0.04$&$16.34\pm0.20$& $6.39\pm0.13$ &$6.49\pm0.13$&$3.24\pm0.09$  \\ \hline \hline
    \end{tabular}
  }
  \label{ST:realdata}
\end{table*}

\vspace{-0.0cm}

\section{Double-tag selection}

The $D^+$ signal candidates are selected from the tracks remaining after the $D^-$ candidates have been formed.
Candidates for $K^\pm$, $\pi^\pm$, $K_{S}^{0}$ and $\pi^0$ are selected using the same criteria as those used in the tag selection.
Candidates for $\eta$ are selected in the same way as $\pi^{0}$, except that the invariant mass of the photon pair is required to be within $(0.115,\,0.150)$\,GeV$/c^{2}$.
For the decays $D^{+} \to K^+ K^- \pi^+ \pi^+ \pi^-$, $D^{+} \to K_S^0 K^+ \pi^+ \pi^- \pi^0$,
and $D^{+} \to K_S^0 K^+ \eta_{\gamma \gamma}$,
we require the signal sides to have 5, 5, and 3 additional charged tracks, respectively.
Additionally, at least two good photons are required for the latter two decays.

To reject background events
from $D^+\to K^+K^-K^0_S\pi^+$ and $D^+\to K^0_SK^+K^0_S\pi^0$ with $K^0_S\to \pi^+\pi^-$ for
the invariant mass of any $\pi^+\pi^-$ combination ($M_{\pi^+\pi^-}$) directly originating from $D$ decays is required to be outside the range $|M_{\pi^+\pi^-}-0.498|<0.020$~GeV/$c^2$.
This requirement corresponds to at least five times its resolution around the $K^0_S$ known mass.

For each selected combination, the energy difference
$\Delta E_{\rm sig} \equiv E_{D^+}-E_{\mathrm{beam}}$ and the beam-constrained mass $M_{\rm BC}^{\rm sig}$ are
calculated, similar to those on the tag side.
If there are multiple combinations,
the one giving the smallest $|\Delta E_{\rm sig}|$ is retained.
In the measurements of hadronic $D^+$ decays, no $M_{\rm BC}^{\rm tag}$ requirement is imposed on the tag side to avoid peaking backgrounds
in the $M_{\rm BC}^{\rm sig}$ distribution.

The $\Delta E_{\rm sig}$ requirements for different signal decays are listed in
Table~\ref{tab:delE_sig}.  These requirements correspond to a region of $3.5\sigma$ around the peak obtained from the fit to the signal MC simulation.

\begin{table}[htbp]
  \centering
  \caption{The $\Delta E_{\rm sig}$ requirements for different
  signal decays. }
  \label{tab:delE_sig}
  \begin{tabular}{lc}
    \hline\hline
    Signal decay                  &
    $\Delta E_{\rm sig}$~(MeV) \\ \hline
    $D^+\to K^{+}K^{-}\pi^+\pi^+\pi^- $
                                  & $(-24,\,22)$            \\
    $D^+\to K^0_SK^+\pi^+\pi^-\pi^0$     &
    $(-31,\,26)$            \\
    $D^+\to K^0_SK^+\eta_{\gamma \gamma}$&
    $(-23,\,22)$            \\ \hline\hline
  \end{tabular}
\end{table}

To extract the signal yields of the $D^+\to K^+K^-\pi^+\pi^+\pi^-$, $D^+\to
K^0_SK^+\pi^+\pi^-\pi^0$, and $D^+\to K^0_SK^+\eta_{\gamma\gamma}$ decays, unbinned maximum likelihood
fits are performed on the $M_{\rm BC}^{\rm sig}$ distributions using individual accepted candidates.
In the fits, the signal shapes are described by the MC-simulated shape convolved with
a Gaussian function with floating parameters to account for the resolution difference between the data and the MC simulation. The background shapes are modeled
by the ARGUS function~\cite{argus}.

\begin{figure*}[htbp]
  \centering
  \hspace{-45pt}
  \includegraphics[width=1.07\linewidth]{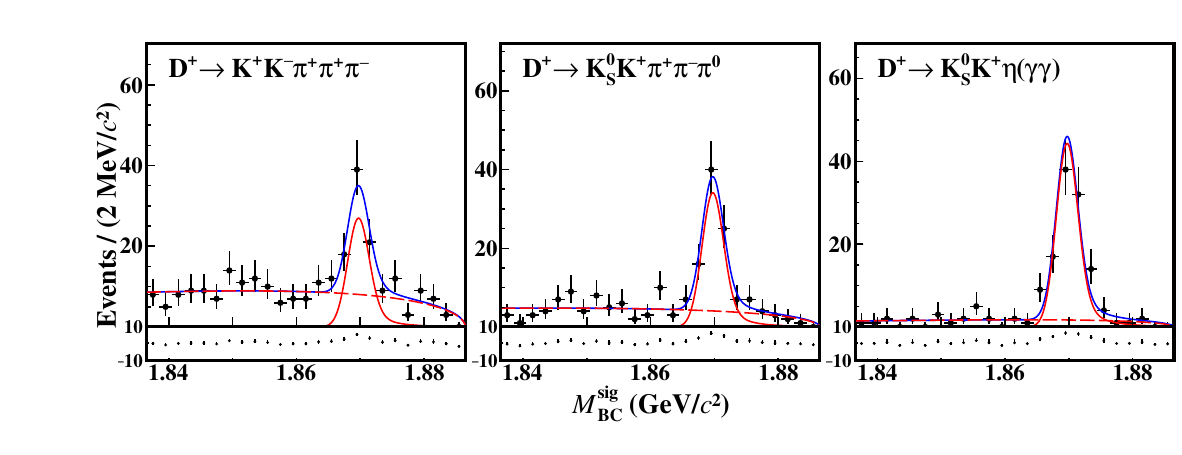}
  \caption{
    The $M_{\rm BC}^{\rm sig}$ distributions for the DT candidate events in
    data, with fit results overlaid.
    The dots with error bars are data, the blue solid curves are the fit
    results, the red solid curves are the fitted signal shapes, and the dashed
  curves are the fitted background shapes.}
  \label{fig:fit2}
\end{figure*}

\begin{figure*}[htbp]
  \centering
  \scalebox{1.14}{ \hspace{-40pt}
    \includegraphics[width=1.\linewidth]{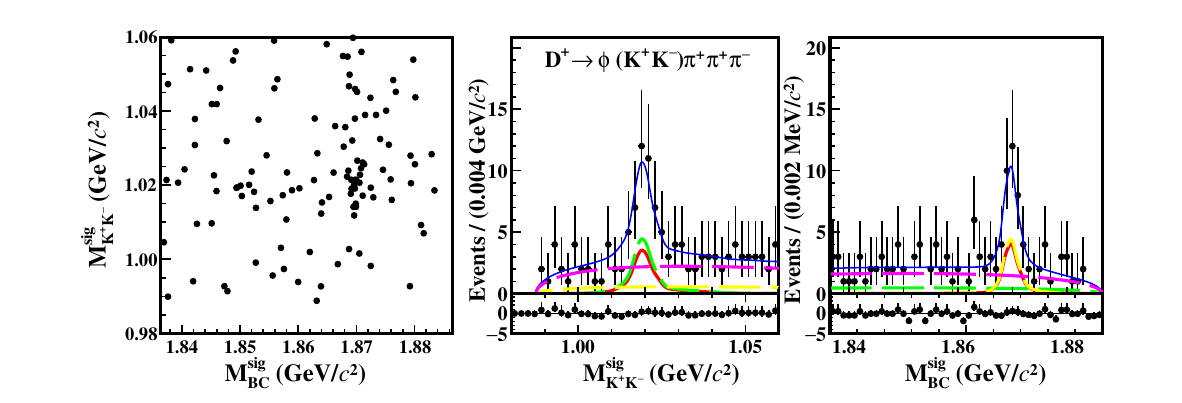}
  }
  \scalebox{1.14}{ \hspace{-40pt}
    \includegraphics[width=1.\linewidth]{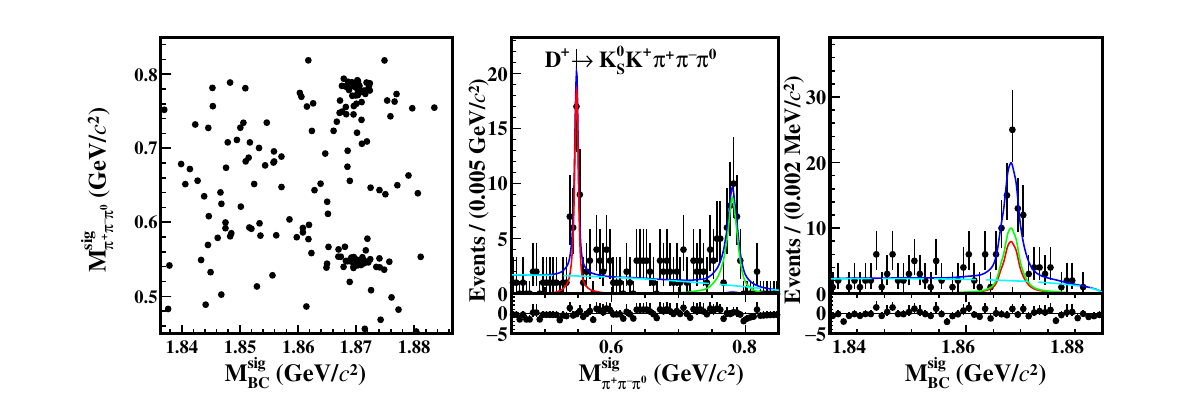}
  }
  \caption{ (Left) The distributions of
    $M_{K^{+}K^{-}/\pi^{+}\pi^{-}\pi^{0}}^{\rm sig}$ versus $M^{\rm sig}_{\rm
    BC}$ of the DT candidate events in data, and the projections of the 2D fits
    to (middle) $M_{K^{+}K^{-}/\pi^{+}\pi^{-}\pi^{0}}^{\rm sig}$ and (right)
    $M_{\rm BC}^{\rm sig}$.  In the middle and right columns, the dots with error
    bars are data, the blue solid curves are the fit results, the dashed curves are the  fitted background shapes, and the solid curves
    in other colors are the  fitted signal shapes.
    In the upper figure, the red solid line represents the $\phi$ signal; in the lower figure,
  the red solid line represents the $\eta$ signal and the green solid line represents the $\omega$ signal.}
  \label{fig:fit}
\end{figure*}

To extract the signal yields of $D^+\to \phi\pi^+\pi^+\pi^-$, $D^+\to
K^0_SK^+\eta_{\pi^+\pi^-\pi^0}$ and $D^+\to K^0_SK^+\omega$,
two-dimensional fits are performed on the distributions of
$M_{K^{+}K^{-}/\pi^{+}\pi^{-}\pi^{0}}^{\rm sig}$ versus $M^{\rm sig}_{\rm BC}$
for the DT candidate events in data.  In the fits, the signal and background
shapes in the $M^{\rm sig}_{\rm BC}$ distributions are described as follows: the
shapes of the $\phi$ and $\eta/\omega$ signals are modeled by the signal MC
shape convolved with a Gaussian function, and the combinatorial
backgrounds are modeled by a reverse-ARGUS function and a second-order polynomial function in the $M_{K^+K^-}$ and $M_{\pi^+\pi^-\pi^0}$ distributions, respectively.
The results of the fits to the accepted DT candidate events in data are shown
in Figs.~\ref{fig:fit2} and~\ref{fig:fit}.

For the decays containing a $K^0_S$ in the final state,
the combinatorial $\pi^+\pi^-$ background in the $K^0_S$ signal region is estimated using
events in the $K^0_S$ sideband region, defined by $0.020<|M_{\pi^+\pi^-}-0.498|<0.044$~GeV/$c^2$.
For signal channels with sideband events present in the data, both the yield and the efficiency have been corrected to account for the influence of the sideband.
These corrections are negligible for all decays.

The statistical significance of the decay $D^+\to \phi \pi^+\pi^+\pi^-$ is $3.1\sigma$,
while the statistical significances of the other signal decays exceed $5.0\sigma$.
They are estimated by comparing the likelihoods of fits including and excluding the signal components, taking into account the change in the number of degrees of freedom.

The yields of the DT events, the efficiencies of detecting the signal decays in
the presence of the ST $D^-$, and the resulting BFs are listed in Table~\ref{tab:result}.
The average BF of $D^+\to K^0_SK^+\eta$ is obtained by re-weighting
the BFs of $D^+\to K^0_SK^+\eta_{\gamma\gamma}$ and $D^+\to K^0_SK^+\eta_{\pi^+\pi^-\pi^0}$,
incorporating the statistical uncertainties and both correlated and uncorrelated systematic uncertainties, as described in Ref.~\cite{corsu}.
Here, the uncertainties arising from the ST yield, $K^{\pm}$ tracking and PID,
$K^0_S$ reconstruction, and the quoted BF of $K^0_S\to \pi^+\pi^-$ are assumed to be correlated,
while those from the other sources are assumed to be uncorrelated. Details of the systematic
uncertainties in the BF measurements are provided below.

\begin{table*}[htbp]
  \centering
  \caption{ The numbers used to calculate the BFs ($\mathcal B_{\rm sig}$),
    where the first uncertainties are statistical and the second systematic.  $N_{\rm DT}$ is the
    number of DT events; $\epsilon_{\rm sig}$ is the efficiency
    of detecting the signal decay in the presence of the ST $D^-$ and does not
  include the sub-decay BFs; $\mathcal B_{\rm weight}$ is obtained by weighting measurements with different $\eta$ decays; and $\mathcal B_{\rm PDG}$ is the previous measurements.}
  \label{tab:result}
  \centering
  \begin{tabular}{lccccc}
    \hline
    \hline
    Signal decay
        & $N_{\rm DT}$        & $\epsilon_{\rm
  sig}~(\%)$ & $~~~\mathcal B_{\rm sig}~(\times10^{-4})$ & $\mathcal B_{\rm weight}~(\times10^{-4})$&$\mathcal B_{\rm
  PDG}~(\times10^{-4})$ \\ \hline
    $D^+\to K^+K^-\pi^+\pi^+\pi^-$           &
    $~61.8\pm10.5$  & $~8.53\pm0.08$
                    & $~~~0.66\pm0.11\pm0.03$          &   ...   &  $2.3\pm1.2$          \\
    $D^+\to \phi\pi^+\pi^+\pi^-$     & $~17.8\pm~6.3$
                                     &                     $~6.09\pm0.07$            &
    $~~~0.54\pm0.19\pm0.02$         &  ...     &      ...     \\ \hline
    $D^+\to K^0_SK^+\pi^+\pi^-\pi^0$         & $~77.5\pm10.5$
                                             & $~4.05\pm0.06$            & $~~~2.51\pm0.34\pm0.14$      &    ...      &     ...     \\ \hline
    $D^+\to K^0_SK^+\eta_{\gamma\gamma}$     &
    $101.5\pm11.2$ &          $15.55\pm0.11$          &
    $~~~2.17\pm0.24\pm0.08$     &    ~~~\multirow{2}*{$2.27\pm0.22\pm0.05$}        &   \multirow{2}*{$1.8\pm0.5$}          \\
    $D^+\to K^0_SK^+\eta_{\pi^+\pi^-\pi^0}$   & $~31.4\pm~6.3$
                                              &                    $~6.17\pm0.07$            &
    $~~~2.90\pm0.59\pm0.15$      &          &        \\  \hline
    $D^+\to K^0_SK^+\omega$ & $~40.4\pm~7.0$
                            & $~2.93\pm0.05$            & $~~~2.02\pm0.35\pm0.10$      & ...         &   ...        \\
                            \hline\hline
  \end{tabular}
\end{table*}

\section{Systematic uncertainties}

With the DT method, the systematic uncertainties associated with the ST selection
cancel and do not affect the BF measurements. The systematic
uncertainties in the selection of the signal side are discussed below.

The uncertainty related to the total ST yield ($N^{\rm tot}_{\rm ST}$) is assigned as 0.3\%,
determined by varying the signal and background
shapes, and floating the parameters of a single Gaussian function in the
fit.

The tracking efficiencies of $(K/\pi)^\pm$ are studied by analyzing DT $D\bar
D$  events, using a control sample where a $(K/\pi)^\pm$ is missing, derived from $D^0 \to K^{-}\pi^{+}$, $D^0 \to K^{-}\pi^{+}\pi^{+}\pi^{-}$, and $D^{+} \to K^{-} \pi^{+}\pi^{+}$ decays.
The systematic
uncertainties in the tracking are assigned as 0.5\% per track.

The PID efficiencies of $(K/\pi)^\pm$ are investigated using the same control samples as in the tracking study.
The systematic
uncertainties in PID are assigned as 0.5\% per track.

The systematic uncertainty in $\pi^{0}$ reconstruction
is assigned as 2.0\% per $\pi^0$ using control samples of $D^0\to K^-\pi^+\pi^0$, $\bar D^0\to K^+\pi^-$ and $\bar D^0\to K^+\pi^-\pi^-\pi^+$.
Due to the limited size of the $\eta_{\gamma\gamma}$ sample, the uncertainty in $\eta_{\gamma\gamma}$ reconstruction is assigned as 2.0\% per $\eta_{\gamma\gamma}$
by referring to the $\pi^{0}$ reconstruction.

The uncertainties of the quoted BFs of $\pi^0\to\gamma\gamma$, $K_{S}^{0}\to \pi^{+}
\pi^{-}$, $\eta\to\gamma\gamma$, $\eta\to\pi^+\pi^-\pi^0$, and
$\omega\to\pi^+\pi^-\pi^0$ are
0.03\%, 0.07\%, 0.5\%, 1.2\%, and 0.8\%, respectively.

The uncertainties due to the $\Delta E$ requirements are assigned by analyzing
the decays $D^{+} \to K^- \pi^+ \pi^+ \pi^+ \pi^-$, $D^{+} \to K^{0}_{S} \pi^+
\pi^+ \pi^- \pi^0$, and $D^{+} \to K^{0}_{S} \pi^+ \eta$.
The differences of the acceptance efficiencies between data and MC simulation are taken as the
systematic uncertainties, which are 0.2\%, 1.6\%, and 0.1\% for $D^{+} \to K^+
K^- \pi^+ \pi^+ \pi^-$, $D^{+} \to K^{0}_{S} K^+ \pi^+ \pi^- \pi^0$, and $D^{+}
\to K^{0}_{S} K^+ \eta$, respectively.

The systematic uncertainties of the mass fit are examined by varying signal and background shapes.
The alternative signal shapes are chosen as the MC truth-matched signal shapes.
The alternative background shapes are obtained by modifying the reverse-ARGUS endpoint
by $\pm0.2$~MeV/$c^2$ and varying the polynomial function by one order.
The largest changes in the re-measured BFs are assigned as the systematic uncertainties.

The $K^0_S$ reconstruction is examined using hadronic
$D\bar D$ events, with $D^0$ or $D^+$ decaying into
$K^0_S\pi^+\pi^-$, $K^0_S\pi^+\pi^-\pi^0$
$K^0_S\pi^0$, $K^0_S\pi^+$, $K^0_S\pi^+\pi^0$, and $K^0_S\pi^+\pi^+\pi^-$.
The systematic uncertainties are assigned as 2.0\% per $K^0_S$.
The systematic uncertainty of  $K_S^0 \to \pi^+\pi^-$ rejection is
studied using the control sample of $D^{+} \to K_{S}^{0} e^+ \nu_{e} $.
The acceptance efficiencies of data and MC simulation are consistent with each other,
and the relevant systematic uncertainty is negligible.

The uncertainty due to the MC sample size is assigned as $(0.5-1.3)\%$ for different signal decays.
The uncertainties due to the mixing of signal MC sample are examined by varying the fractions by $\pm 1 \sigma$. The changes of
the efficiencies are assigned as the systematic uncertainties.

The details of the systematic uncertainties in the BF measurements are listed in Table~\ref{tab:relsysuncertainties}.

\begin{table*}[htbp]
  \centering
  \caption{ Relative systematic uncertainties
  (\%) in the BF measurements.  }
  \label{tab:relsysuncertainties}
  \centering
  \begin{tabular}{ccccccc} \hline\hline Signal decay
      &$K^+K^-\pi^+\pi^+\pi^-$&$K^0_SK^+\pi^+\pi^-\pi^0$&$K^0_SK^+\eta_{\gamma\gamma}$&$\phi\pi^+\pi^+\pi^-$&$K^0_SK^+\eta_{\pi^+\pi^-\pi^0}$&$K^0_SK^+\omega$\\
      \hline
    $N^{\rm tot}_{\rm ST}$  & 0.3  & 0.3 &0.3 & 0.3 & 0.3 & 0.3 \\
    $(K/\pi)^{\pm}$ tracking  & 2.5  & 2.5 & 1.5 & 2.5 & 2.5 & 2.5 \\
    $(K/\pi)^{\pm}$ PID & 2.5  & 1.5 & 0.5 & 2.5 & 1.5 & 1.5 \\
    $\pi^0$ or $\eta_{\gamma\gamma}$ reconstruction       &  ...   & 2.0   & 2.0   &  ...  & 2.0 & 2.0 \\
    $\mathcal B_{\rm sub-decay}$  &  ...   & 0.1   & 0.5 & 1.0 & 1.2 & 0.8 \\
    $\Delta E_{\rm sig}$ cut  & 0.2  & 1.6 & 0.0 & 0.2 & 1.6 & 1.6 \\
    Mass fit  & 0.9  & 1.2 & 1.9 & 2.2 & 2.4 & 2.4 \\
    $K_{S}^{0}$ reconstruction &  ...   & 2.0 & 2.0 &  ...  & 2.0 & 2.0\\
    MC sample size & 0.8  & 1.1 & 0.5 & 0.9 & 0.8 & 1.3 \\
    MC model & 2.6  & 3.2 & ...& ...& ...& ... \\
    \hline
    Total & 4.3 & 5.6 & 3.8 & 4.3 & 5.2 & 5.2 \\
    \hline\hline
  \end{tabular}
\end{table*}

\section{Summary}

In summary, by analyzing 20.3~fb$^{-1}$ of
$e^+e^-$ collision data collected with the BESIII detector at $\sqrt s=3.773$~GeV,
the absolute BFs of
$D^+\to K^+K^-\pi^+\pi^+\pi^-$, $D^+\to K^0_SK^+\eta$,
$D^+\to \phi\pi^+\pi^+\pi^-$,
$D^+\to K^0_SK^+\pi^+\pi^-\pi^0$, and $D^+\to K^0_SK^+\omega$ have been measured to be
$(0.66\pm0.11\pm0.03)\times 10^{-4}$, $(2.27\pm0.22\pm0.05)\times 10^{-4}$, $(0.54\pm0.19\pm0.02)\times 10^{-4}$,
$(2.51\pm0.34\pm0.14)\times 10^{-4}$, and
$(2.02\pm0.35\pm0.10)\times 10^{-4}$, respectively.
The BFs of the latter three decays are measured for the first time,
while those of the former two decays are measured with improved precision.
Our results enrich the knowledge of the $D\to K \bar{K} \pi\pi\pi$ hadronic decays.
Larger data in the future will be helpful to further study the substructures in these decays.

\section{Acknowledgement}

The BESIII Collaboration thanks the staff of BEPCII and the IHEP computing center for their strong support. This work is supported in part by National Key R\&D Program of China under Contracts Nos. 2023YFA1606000; National Natural Science Foundation of China (NSFC) under Contracts Nos. 11635010, 11735014, 11935015, 11935016, 11935018, 12025502, 12035009, 12035013, 12061131003, 12192260, 12192261, 12192262, 12192263, 12192264, 12192265, 12221005, 12225509, 12235017, 12361141819; the Chinese Academy of Sciences (CAS) Large-Scale Scientific Facility Program; the CAS Center for Excellence in Particle Physics (CCEPP); Joint Large-Scale Scientific Facility Funds of the NSFC and CAS under Contract No. U1832207; CAS under Contract No. YSBR-101; 100 Talents Program of CAS; The Institute of Nuclear and Particle Physics (INPAC) and Shanghai Key Laboratory for Particle Physics and Cosmology; Agencia Nacional de Investigación y Desarrollo de Chile (ANID), Chile under Contract No. ANID PIA/APOYO AFB230003; German Research Foundation DFG under Contract No. FOR5327; Istituto Nazionale di Fisica Nucleare, Italy; Knut and Alice Wallenberg Foundation under Contracts Nos. 2021.0174, 2021.0299; Ministry of Development of Turkey under Contract No. DPT2006K-120470; National Research Foundation of Korea under Contract No. NRF-2022R1A2C1092335; National Science and Technology fund of Mongolia; National Science Research and Innovation Fund (NSRF) via the Program Management Unit for Human Resources \& Institutional Development, Research and Innovation of Thailand under Contract No. B50G670107; Polish National Science Centre under Contract No. 2019/35/O/ST2/02907; Swedish Research Council under Contract No. 2019.04595; The Swedish Foundation for International Cooperation in Research and Higher Education under Contract No. CH2018-7756; U. S. Department of Energy under Contract No. DE-FG02-05ER41374.


\begin{thebibliography}{0}%
\makeatletter
\providecommand \@ifxundefined [1]{%
 \@ifx{#1\undefined}
}%
\providecommand \@ifnum [1]{%
 \ifnum #1\expandafter \@firstoftwo
 \else \expandafter \@secondoftwo
 \fi
}%
\providecommand \@ifx [1]{%
 \ifx #1\expandafter \@firstoftwo
 \else \expandafter \@secondoftwo
 \fi
}%
\providecommand \natexlab [1]{#1}%
\providecommand \enquote  [1]{``#1''}%
\providecommand \bibnamefont  [1]{#1}%
\providecommand \bibfnamefont [1]{#1}%
\providecommand \citenamefont [1]{#1}%
\providecommand \href@noop [0]{\@secondoftwo}%
\providecommand \href [0]{\begingroup \@sanitize@url \@href}%
\providecommand \@href[1]{\@@startlink{#1}\@@href}%
\providecommand \@@href[1]{\endgroup#1\@@endlink}%
\providecommand \@sanitize@url [0]{\catcode `\\12\catcode `\$12\catcode
  `\&12\catcode `\#12\catcode `\^12\catcode `\_12\catcode `\%12\relax}%
\providecommand \@@startlink[1]{}%
\providecommand \@@endlink[0]{}%
\providecommand \url  [0]{\begingroup\@sanitize@url \@url }%
\providecommand \@url [1]{\endgroup\@href {#1}{\urlprefix }}%
\providecommand \urlprefix  [0]{URL }%
\providecommand \Eprint [0]{\href }%
\providecommand \doibase [0]{https://doi.org/}%
\providecommand \selectlanguage [0]{\@gobble}%
\providecommand \bibinfo  [0]{\@secondoftwo}%
\providecommand \bibfield  [0]{\@secondoftwo}%
\providecommand \translation [1]{[#1]}%
\providecommand \BibitemOpen [0]{}%
\providecommand \bibitemStop [0]{}%
\providecommand \bibitemNoStop [0]{.\EOS\space}%
\providecommand \EOS [0]{\spacefactor3000\relax}%
\providecommand \BibitemShut  [1]{\csname bibitem#1\endcsname}%
\let\auto@bib@innerbib\@empty
\end{thebibliography}%


\begin{thebibliography}{99}

  \bibitem{Li:2021iwf}
  H.~B.~Li and X.~R.~Lyu,
  \href{https://arxiv.org/abs/2103.00908}{Natl. Sci. Rev. \textbf{8} (2021) no.11, nwab181.}

  \bibitem{hflav2018} Y. Amhis {\it et al.} (Heavy Flavor Averaging Group),
  \href{https://arxiv.org/abs/2411.18639}{arXiv:2411.18639.}

  \bibitem{lhcbnote} R. Aaij {\it et al.} (LHCb Collaboration),
  \href{http://cds.cern.ch/record/2223391?ln=zh_CN}{Synergy of BESIII and
  LHCb physics programmes, LHCb-PUB-2016-025.}

  \bibitem{ref::pdg2024} S. Navas {\it et al.} (Particle Data Group),
  \href{https://doi.org/10.1103/PhysRevD.110.030001}{{Phys. Rev. D} {\bf
  110}, 030001 (2024)}.

  \bibitem{kkppp} J. M. Link {\it et al.} (FOCUS Collaboration),
  \href{https://doi.org/10.1016/S0370-2693(03)00494-5}{Phys. Lett. B {\bf
  561}, 225-232 (2003).}

  \bibitem{kketa} M. Ablikim {\it et al}. (BESIII Collaboration),
  \href{https://doi.org/10.1103/PhysRevLett.124.241803}{Phys. Rev. Lett. {\bf
  124}, 241803 (2020).}

  \bibitem{BESIII:2024lbn} M. Ablikim {\it et al}. (BESIII
  Collaboration),\\
  \href{https://arxiv.org/abs/2406.05827}{arXiv:2406.05827}.

  \bibitem{Ablikim:2009aa} M. Ablikim {\it et al}. (BESIII
  Collaboration),~\href{https://linkinghub.elsevier.com/retrieve/pii/S0168900209023870}{{Nucl.\
  Instrum.\ Meth.\ A} {\bf 614}, 345 (2010)}.

  \bibitem{Yu:IPAC2016-TUYA01} C.~H.~Yu {\it et al.},
  \href{doi:10.18429/JACoW-IPAC2016-TUYA01}{Proceedings of IPAC2016, Busan,
  Korea, 2016.}

  \bibitem{Ablikim:2019hff} M. Ablikim {\it et al}. (BESIII Collaboration),
  \href{https://iopscience.iop.org/article/10.1088/1674-1137/44/4/040001}{{Chin.
  Phys. C} {\bf 44}, 040001 (2020)}.	

  \bibitem{EcmsMea} J. Lu, Y. Xiao, and X, Ji,
  ~\href{https://link.springer.com/article/10.1007/s41605-020-00188-8#citeas}{{Radiat.
  Detect. Technol. Methods} {\bf 4}, 337 (2020)}.

  \bibitem{EventFilter} J.~W.~Zhang {\it et al.},
  ~\href{https://link.springer.com/article/10.1007/s41605-022-00331-7}{{Radiat.
  Detect. Technol. Methods} {\bf 6}, 289 (2022)}.

  \bibitem{detvis} K.~X.~Huang {\it et al.},
  \href{https://link.springer.com/article/10.1007/s41365-022-01133-8}{{Nucl.\
  Sci.\ Tech} {\bf 33}, 142 (2022).}

  \bibitem{etof} X.~Li {\it et al.},
  ~\href{https://link.springer.com/article/10.1007/s41605-017-0012-4}{Radiat.
  Detect. Technol. Methods {\bf 1}, 13 (2017)}; ~Y.~X.~Guo {\it et al.},
  ~\href{https://link.springer.com/article/10.1007/s41605-017-0012-4}{{Radiat.
  Detect. Technol. Methods} {\bf 1}, 15 (2017)}; ~P.~Cao {\it et al.},
  \href{https://www.sciencedirect.com/science/article/pii/S0168900219314068?via\%3Dihub}{{Nucl.\
  Instrum.\ Meth.\ A} {\bf 953}, 163053 (2020)}.

  \bibitem{geant4} S. Agostinelli {\it et al.} (GEANT4 Collaboration),
  \href{https://www.sciencedirect.com/science/article/pii/S0168900203013688?via\%3Dihub}{{Nucl.
  Instrum. Meth. A} {\bf 506}, 250 (2003)}.

  \bibitem{kkmc} S. Jadach, B. F. L. Ward, and Z. Was,
  \href{https://www.sciencedirect.com/science/article/pii/S0010465500000485?via\%3Dihub}{
  {Comp. Phys. Commu} {\bf 130}, 260 (2000)};
  ~\href{https://journals.aps.org/prd/abstract/10.1103/PhysRevD.63.113009}{{Phys.
  Rev. D} {\bf 63}, 113009 (2001)}.

  \bibitem{evtgen} D.~J.~Lange,
  \href{https://doi.org/10.1016/S0168-9002(01)00089-4} {{Nucl. Instrum. Meth.
  A} {\bf 462}, 152 (2001)}; R.~G.~Ping,
  \href{https://doi.org/10.1088/1674-1137/32/8/001}{{Chin. Phys. C} {\bf 32},
  599 (2008).}

  \bibitem{lundcharm} J. C. Chen, G. S. Huang, X. R. Qi, D. H. Zhang, and Y. S.
  Zhu,
  \href{https://journals.aps.org/prd/abstract/10.1103/PhysRevD.62.034003}{{Phys.
  Rev. D} {\bf 62}, 034003 (2000)}.

  \bibitem{photos} E.~Richter-Was,
  ~\href{https://doi.org/10.1016/0370-2693(93)90062-M`}{{Phys. Lett. B} {\bf
  303}, 163 (1993)}.

  \bibitem{FOCUS:2001omf} J.~M.~Link {\it et al.} (FOCUS Collaboration),
  \href{https://doi.org/10.1103/PhysRevLett.87.162001}{Phys. Rev. Lett. {\bf
  87}, 162001 (2001).}

  \bibitem{MARKIII1} R. M. Baltrusaitis {\it et al.} (MARKIII Collaboration),
  \href{https://doi.org/10.1103/PhysRevLett.56.2140}{Phys. Rev. Lett. {\bf
  56}, 2140 (1986).}

  \bibitem{MARKIII2} J. Adler {\it et al.} (MARKIII Collaboration),
  \href{https://doi.org/10.1103/PhysRevLett.60.89}{Phys. Rev. Lett. {\bf 60},
  89 (1988).}

  \bibitem{emcshower} M. Ablikim {\it et al}. (BESIII
  Collaboration),
  \href{https://doi.org/10.1016/j.nima.2009.12.050}{{Nucl. Instrum. Metho. A} {\bf 614}, 345 (2010)};
  \href{http://dx.doi.org/10.1088/1674-1137/44/4/040001}{{Chin. Phys. C} {\bf 44}, 040001 (2020).}

  \bibitem{argus} H. Albrecht {\it et al.} (ARGUS Collaboration),
  \href{https://doi.org/10.1016/0370-2693(90)91293-K}{{Phys. Lett. B} {\bf
  241}, 278 (1990).}

  \bibitem{corsu} M. Ablikim {\it et al}. (BESIII
  Collaboration),
  \href{https://journals.aps.org/prd/abstract/10.1103/PhysRevD.108.092003}{{Phys.
  Rev. D} {\bf 108}, 092003 (2023)}.

\end{thebibliography}
\end{document}